\documentclass[a4paper,12pt]{article}

\usepackage{amsmath}
\usepackage{amssymb}
\usepackage{dsfont}
\usepackage{bbm}
\usepackage{epsf}
\usepackage{epsfig}
\usepackage{graphicx}
\usepackage[english]{babel}
\usepackage[latin1]{inputenc}
\usepackage{rotating}
\usepackage{subfigure}
\usepackage{color}
\usepackage{slashed}
\usepackage{wrapfig}

\usepackage{txfonts}
\newcommand{\deltat}{a_{\text{t}}}
\newcommand{\trace}{\text{Tr} }

\newcommand{\epsf}{\epsilon^{1/4}}
\newcommand{\epsmf}{\epsilon^{-1/4}}

\newcommand{\mywidth}{8.5cm}
\renewcommand{\vec}[1]{{\bf #1}}

\begin{document}

\begin{center}
\begin{Large}
\textbf{Bottom-up isotropization in classical-statistical lattice gauge theory} \\
\end{Large}

\medskip

J\"urgen Berges, Sebastian Scheffler, and D\'enes Sexty \\

\textit{Institute for Nuclear Physics, Darmstadt University of Technology, Schlossgartenstr. 9, 64285 Darmstadt, Germany}\\
\end{center}

\begin{abstract}
We compute nonequilibrium dynamics for classical-statistical SU(2) pure gauge theory on a lattice. We consider anisotropic initial conditions with high occupation numbers in the transverse plane on a characteristic scale $\sim Q_s$. This is used to investigate the very early stages of the thermalization process in the context of heavy-ion collisions. We find Weibel or "primary" instabilities with growth rates similar to those obtained from previous treatments employing anisotropic distributions of hard modes (particles) in the weak coupling limit. We observe "secondary" growth rates for higher-momentum modes reaching substantially larger values and we analyse them in terms of resummed loop diagrams beyond the hard-loop approximation. We find that a coarse grained pressure isotropizes "bottom-up" with a characteristic inverse rate of $\gamma^{-1} \simeq 1 - 2$fm/c for coarse graining momentum scales of $p \lesssim 1$ GeV choosing an initial energy density for RHIC of $\epsilon = 30 \text{GeV/fm}^3$. The nonequilibrium spatial Wilson loop is found to exhibit an area law and to become isotropic on a similar time scale. 
\end{abstract}

\section{Introduction}\label{Introduction}
The theoretical understanding of the thermalization process in collision experiments of
heavy nuclei at RHIC (BNL), the LHC (CERN) or the planned FAIR facility at GSI is crucial
for a successful outcome of the heavy-ion program. Available data from RHIC reveals
remarkable properties such as rapid apparent thermalization with robust collective
phenomena. There seems to be a good description of a wide range of data using
hydrodynamical models, provided one assumes a fast approach to local thermal equilibrium
in less than about $1$fm/c~\cite{Heinz:2004pj}. The understanding of these extremely short time
scales from QCD provides a challenge for theory. It has been pointed out that complete
local thermal equilibrium may not be necessary for the application of hydrodynamics, and
a prethermalization of the equation of state could be sufficient~\cite{Berges:2005ai}.
Filamentation or so-called Weibel instabilities in anisotropic plasmas have been
identified as the parametrically fastest processes which may help to explain a rapid
isotropization of the equation of state on time scales much shorter than characteristic thermal
equilibration times~\cite{Mrowczynski:1988dz, isotropization, WongYangMills, Romatschke:2006nk}.

During the past years several groups investigated plasma instabilities in non-equilibrium
gauge theories within different frameworks. Extensive studies have been performed within
the "hard-loop approximation", which is equivalent to the collisionless Vlasov
theory~\cite{Mrowczynski:1988dz}, or solving the effective Wong-Yang-Mills equations~\cite{WongYangMills}. Another
important development concerns the solution of classical Yang-Mills fields in the
presence of static colored source terms using the framework of the color glass condensate
effective theory in expanding geometries~\cite{Romatschke:2006nk}. Also transport or kinetic equations in
the spirit of earlier "bottom-up thermalization" scenarios~\cite{bottom-up-thermalization}, i.e. assuming that no
plasma instabilities occur, were performed with remarkably short thermalization
times~\cite{Xu:2007aa}.

Typically, these studies rely on a sufficiently large separation of scales between
suitably defined soft and hard momentum scales. This is applicable only at small
characteristic running gauge coupling and limits many results to the weak coupling
regime. One would like to know what happens in the absence of a clear separation of
scales between hard and soft modes or when the initial fields have large amplitudes which
is seemingly the case in present day experiments. Such nonperturbative studies may be
based on simulations of lattice gauge theory using real-time stochastic quantization
techniques~\cite{stochquant}, or classical-statistical methods~\cite{Aarts:2000wi} in the presence of sufficiently
large occupation numbers. Classical-statistical simulations on a lattice have been
extremely successful in the past in describing the nonlinear scalar inflaton dynamics in
the early universe in the presence of parametric resonance or spinodal instabilities after
inflation~\cite{preheating,Podolsky:2005bw,Aarts:2001yn}. These instabilities lead to exponential growth of occupation numbers
in long wavelength modes on time scales much shorter than the asymptotic thermal
equilibration time. Characteristic far-from-equilibrium phenomena, such as an early
prethermalization of the equation of state, have been quantitatively studied in that context~\cite{Berges:2005ai,Podolsky:2005bw}. Comparisons with calculations including quantum corrections demonstrate
the accurate description of nonequilibrium quantum field dynamics by classical-statistical
methods for not too late times, i.e.\ before the thermalization characterized by
Bose-Einstein statistics sets in~\cite{Aarts:2001yn}.

In this paper we compute nonequilibrium dynamics for classical-statistical SU(2) pure
gauge theory on a lattice. The theory can be computed without further approximations
using standard techniques by numerical integration and Monte Carlo sampling over
initial conditions for given normalized initial probability functional~\cite{Berges:2004yj}. We consider
anisotropic initial conditions with large occupation numbers in the transverse plane with
the characteristic scale $\Delta$ that may be identified with the saturation scale $\sim
Q_s$~\cite{Gyulassy:2004zy}. Its value is varied in a range $0.5 \lesssim
\Delta/\epsilon^{1/4} \lesssim 2.5$, where $\epsilon$ denotes the initial energy density.
This is used to investigate the very early stages of the thermalization process
(isotropization, prethermalization) such that a static geometry, i.e.\ no expansion may
be considered. We find Weibel or "primary" instabilities with characteristic inverse growth
rates similar to those obtained from previous treatments employing anisotropic
distributions of hard modes (particles) in the weak coupling limit. We observe "secondary"
growth rates for higher-momentum modes reaching substantially larger values. The secondaries are
fluctuation induced and we analyze them in terms of resummed loop diagrams
beyond the hard-loop approximation. We find that a coarse grained pressure isotropizes "bottom-up"
with a characteristic inverse growth rate of $\gamma^{-1} \simeq 1 - 2$fm/c for coarse
graining momentum scales of $p \lesssim 1$ GeV choosing a value for RHIC of
$\epsilon = 30 \text{GeV/fm}^3$. We finally analyze the nonequilibrium
spatial Wilson loop which is found to exhibit an area law. This is used to compute the
time evolution of a "spatial string tension" which is shown to be non-vanishing and
isotropic on a similar characteristic time scale as the low-momentum coarse-grained pressure.

The paper is organized as follows. In section \ref{sec:classical-statistical}, we outline our approach and the numerical implementation. Subsequently, we analyze the characteristic time scales associated to primary and secondary growth rates in section \ref{Results}. Section \ref{sec:bottom-up-isotropization} discusses isotropization of a coarse grained pressure and section \ref{wilson-loop-section} computes nonequilibrium dynamics of large Wilson loops. We conclude with a discussion in section \ref{Discussion} and verify volume and cutoff independence of our results in an appendix.  

\section{Classical-statistical gauge field theory on a lattice}
\label{sec:classical-statistical}

We consider the Wilsonian lattice action for SU(N) gauge theory in Minkowski space-time
\begin{eqnarray}\label{LatticeAction}
S[U] &=& - \beta_0 \sum_{x} \sum_i \left\{ \frac{1}{2 {\rm Tr}
\mathds{1}} \left( {\rm Tr}\, U_{x,0i} + {\rm Tr}\, U_{x,0i}^{\dagger}
\right) - 1 \right\}
\nonumber\\
&& + \beta_s \sum_{x} \sum_{\genfrac{}{}{0pt}{1}{i,j}{i<j}} \left\{ \frac{1}{2 {\rm
Tr} \mathds{1}} \left( {\rm Tr}\, U_{x,ij} + {\rm Tr}\, U_{x,ij}^{\dagger}
\right) - 1 \right\} \, .
\end{eqnarray}
It is described in terms of the plaquette variable
\begin{equation}
U_{x,\mu\nu} \equiv U_{x,\mu} U_{x+\hat\mu,\nu}
U^{\dagger}_{x+\hat\nu,\mu} U^{\dagger}_{x,\nu} \label{eq:plaq}\, ,
\end{equation}
where $U_{x,\nu\mu}^{\dagger}=U_{x,\mu\nu}\,$. Here $U_{x,\mu}$ is the
parallel transporter associated with the link from the neighboring
lattice point $x+\hat{\mu}$ to the point $x$ in the direction of
the lattice axis $\mu = 0,1,2,3$ with $U_{x,\mu} = U^{\dagger}_{x+
\hat\mu,-\mu}\,$. The parameters 
\begin{equation}
\beta_0 \equiv \frac{2 \gamma {\rm Tr} \mathds{1}}{g_0^2} \,\, ,
\quad \beta_s \equiv \frac{2 {\rm Tr} \mathds{1}}{g_s^2 \gamma} \, ,
\label{eq:ganisoM}
\end{equation}
contain the anisotropy parameter $\gamma \equiv a_s/a_t$, where $a_s$ denotes the spatial and $a_t$ the temporal lattice spacings, and we will consider $g_0 = g_s = g$ as the coupling constant of the lattice theory.              
             
In the following we specify to SU(2) as the gauge group. The dynamics is solved in temporal axial gauge, i.\ e.\ $U_{x,0} = \mathds{1}$ and the numerical treatment is similar as in Ref.~\cite{Ambjorn:1990pu}. Varying the action \eqref{LatticeAction} w.r.t.\ the spatial link variables $U_{x, j}$ yields the leapfrog-type equations of motion
\begin{equation}\label{EOM}
\begin{split}
E^b_j(t + \deltat, \vec{x}) \;  = \; &  E^b_j(t, \vec{x}) + \, \frac{i}{2}\, \frac{1}{\gamma^2 a_s a_t g} \, \sum_{k} \, \Bigl\{ \trace \, \bigl(\sigma^b U_{x, j} U_{(x +\hat{j}), k} U^{\dagger}_{( x + \hat{k}), j} U^{\dagger}_{x, k} \bigr) \\ 
& + \trace \, \bigl( \sigma^b U_{x,j} U^{\dagger}_{(x + \hat{j} - \hat{k}) , k} U^{\dagger}_{(x - \hat{k}), j} U_{( x  - \hat{k}) ,k} \bigr) \Bigr\}  \, ,
\end{split}
\end{equation}
where $\sigma^a$ ($a=1,2,3$) are the Pauli matrices. The electric fields are given by~\cite{Kogut:1974ag} 
\begin{equation}\label{DefElectricField}
E^b_j(x) = - \frac{i}{2} \, \frac{1}{a_s \deltat g}  \, \trace \,  \bigl( U_{(x + \hat{t} ), j} U^{\dagger}_{x, j}\sigma^b  ) \,  =  \, - \frac{i}{2} \, \frac{1}{a_s \deltat g} \, \trace \, \bigl( U_{x, 0j} \sigma^b \bigr) \, .
\end{equation}
Thus $E^b_j(t + \deltat, \vec{x}) $ can be computed from the electric fields and the spatial link variables at time $t$. Then the link variables at $t + \deltat$ are given by
\begin{equation}\label{EOM2}
U_{(x+ \hat{t}), j} = \left\{ \, \Bigl( 1 - (a_s a_t g)^2 \, \sum_b \, E^b_j(t + a_t, \vec{x})^2 \Bigr)^{1/2} \, \mathds{1} \, + \, i a_s a_t g E_j^b(t + a_t, \vec{x}) \sigma^b  \right\} U_{x, j} \, .
\end{equation}
Varying the action \eqref{LatticeAction} w.r.t.\ to a temporal link gives the Gauss constraint 
\begin{equation}\label{GaussLaw}
\sum_{j=1}^3 \, \bigl[ E_j^b(x) - U^{\dagger}_{x -\hat{j},j} \, E_j^b(x - \hat{j} \, ) U_{x -\hat{j},j} \, \bigr] =  0 \, . 
 \end{equation}
The coupling $g$ can be scaled out of the equations of motion and we will set $g = 1$ for the simulations. While $g$ is irrelevant for the classical dynamics it appears when calculating observables such as pressure as described below. 
The relation between the link variables and $A_{j}$ is given by $U_{j}(x) = e^{i g_s a_s A_j(x)}$ where $A_j(x) = A_j^a(x) \sigma^a$ and, accordingly, we use
\begin{equation}\label{compute-gauge-field}
A_j^b(t, \vec{x}) \simeq -\frac{i}{2 a_s g} \, \trace \, \bigl( \sigma^b U_j(t, \vec{x}) \bigr)  \; .
\end{equation}
We will often consider the Fourier transform w.r.t.\ the
spatial coordinates, i.e.\ $A^b_j(t, \vec{p})$.

The initial conditions are given in terms of a normalized initial Gaussian probability functional $P[A(t=0) , E(t=0)]$ from which correlation functions, such as the two-point function, are obtained as 
\begin{equation}\label{eq:exp}
\langle \, A^a_i(x) A^b_j(y) \rangle \, = \, \int {\cal D} A(0) {\cal D} E(0) \, P[A(0), E(0)] \, A^a_i(x) A^b_j(y) \,,
\end{equation}
where the measure indicates integration over classical phase space. The initial Gaussian probability functional is chosen such that
\begin{equation}\label{InitCond-1}
\langle \, \vert \, A^b_j(t=0,\vec{p}) \, \vert^2 \, \rangle \, = \, \frac{C}{(2 \pi)^{3/2} \Delta^2 \Delta_z} \, \exp \Bigl\{ -\frac{ p_x^2 + p_y^2 }{2 \Delta^2} - \frac{p_z^2}{2 \Delta_z^2} \Bigr\} \, , 
\end{equation}
and we consider $E^a_j = - \dot{A_j^a} = 0 $ at $t=0$ fulfilling the Gauss constraint~\eqref{GaussLaw}, which ensures that the latter is respected at all times. In practice, we compute expectation values by solving the equations of motion~\eqref{EOM} and~\eqref{EOM2} for a set of initial configurations sampled according to Eq.~\eqref{InitCond-1} and then average over the results from the individual runs. The margins of errors that we quote are the statistical errors. 
The widths appearing in Eq.~\eqref{InitCond-1} are typically chosen such that $\Delta \gg \Delta_z $ and $\Delta_z $ is set to a sufficiently small value to have an effectively $\delta(p_z)$- like distribution on the lattice. We also discuss the effects of broadening the initial longitudinal distribution. Here $\Delta$ determines the typical transverse momenta of the gluons and may be associated with the saturation scale of order GeV. The constant $C$ appearing in ~\eqref{InitCond-1} is adjusted to obtain a given energy density. For numerical reasons, we initialize all Fourier coefficients whose absolute value is less than a small lower bound with a little noise term. We have checked that our results are insensitive to the amplitude of the noise. 

The energy density in lattice units is determined by the Hamiltonian \cite{Kogut:1974ag} corresponding to~\eqref{LatticeAction} which reads 
\begin{equation}\label{DefEnergyDensity}
\hat{\epsilon}(t, \vec{x}) \, \equiv \,  \frac{1}{4 g^2} \, \left\{ \frac{a_s^2}{\deltat^2} \, \sum_j \, \bigl[ 2 -  \trace \, U_{x,0j} \bigr] + \sum_{j < k} \, \bigl[ 2 - \trace \, U_{x, jk} \bigr] \, \right\}  \, .
\end{equation}
For the conversion to physical units we will fix the lattice spacing $a_s$ from the relation between the physical energy density and its lattice analogue according to
\begin{equation}\label{eps-epshat}
\epsilon = \hat{\epsilon} \cdot a_s^{-4} \; .
\end{equation}
If $g$ is taken to be different from one then the lattice spacing is altered by a factor $1 / \sqrt{g}$ , which follows from Eq.~\eqref{DefEnergyDensity}, and the values for $a$ will not be altered significantly as long as $g \sim \mathcal{O}(1)$.  

We have checked that the results, which are presented in the following, are insensitive to the employed volume and lattice spacings. Details are given in an appendix. 

\section{Characteristic early time scales}\label{Results}

Instabilities provide the parametrically fastest nonequilibrium processes, which may help to understand the apparent success of hydrodynamics in describing RHIC data at times of less than about 1 fm/c. A crucial assumption for the use of ideal hydrodynamic models
is that the stress tensor, in the local rest frame at some point in the system, is nearly diagonal, $T_{ij} \sim \delta_{ij}$, with some equation of state relating the pressure to the energy density. This is just a statement of isotropy which may be faster achieved than complete local thermal equilibrium~\cite{isotropization}. In particular, hydrodynamics is seen to provide a description of the data only for not too high characteristic momenta of order GeV~\cite{Heinz:2004pj}. Therefore, an isotropic coarse grained pressure, which does not take into account anisotropies on distance scales smaller than some characteristic inverse coarse graining momentum may be sufficient. Our following results seem to support this possibility. 
Here we concentrate on the very early stages of the dynamics, which is supposed to be quantitative only for sufficiently fast processes such that the expansion of the system can be neglected. We also emphasize that studying the late-time dynamics is beyond any classical-statistical approximation, which cannot explain the approach to thermal equilibrium characterized by Bose-Einstein distributions. For sufficiently short times and large range of highly occupied momentum modes, however, it is expected to provide a quantitative description of the nonperturbative dynamics~\cite{Aarts:2001yn}.

\subsection{Primary and secondary growth rates}\label{sec:growth-rates}

\begin{figure}[t!]
\begin{center}
\epsfig{file=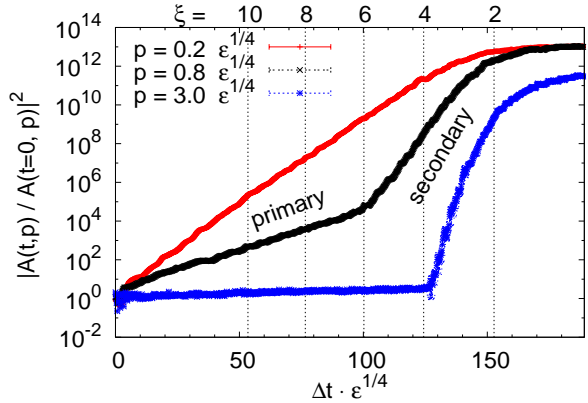, width = \mywidth }
\caption{(Color online) Fourier coefficients of the squared modulus of the gauge field versus time for three different momenta parallel to the z-axis. The low-momentum modes exhibit exponential growth from the beginning. This is followed by a secondary stage that sets in later but with significantly larger growth rates for modes. The curves correspond to the momenta in the legend from top to bottom in the same order.} 
\label{A-vs-t}
\end{center}
\end{figure}
\begin{figure}[t!]
\begin{center}
\epsfig{file=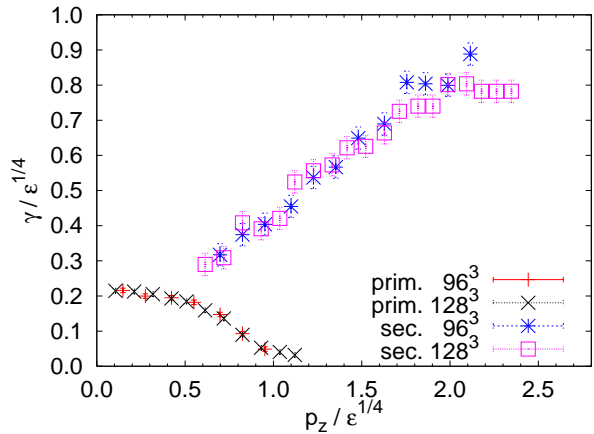, width = \mywidth }
\caption{Primary and secondary growth rates for $| A(t, \vec{p}) |^2$ as a function of $p_z$  as measured on 96$^\text{3}$- and 128$^{3}$- lattices using the same lattice spacing.}
\label{fig-growth-rates}
\end{center}
\end{figure}
In Fig.~\ref{A-vs-t} we plot the color-averaged squared modulus of three different Fourier coefficients of the gauge field modes $A^b_x(t, \vec{p})$ as a function of the difference $\Delta t = t - t_0$ to the initial time $t_0$, normalized by the corresponding field values at $t=t_0$.~\footnote{The data is an average of 100 runs on a 64$^3$- lattice.} All values are given in appropriate units of the initial energy density $\epsilon$. Here $| A(t, \vec{p}) |^2$ may be associated to particle numbers~\cite{smit}. The momenta $\vec{p}$ are chosen parallel to the axis of anisotropy, i.e.\ the z-axis. These modes exhibit the fastest changes in qualitative agreement with analytical results in the weak coupling limit for the growth of Weibel instabilities in non-Abelian gauge theories~\cite{Mrowczynski:1988dz}. The plotted low-momentum modes clearly show exponential growth starting at the very beginning of the simulation. In contrast to these "primary" instabilities operating at low momenta, one observes from Fig.~\ref{A-vs-t} that momentum modes at sufficiently high momenta do not grow initially. The higher wave number modes typically exhibit exponential growth at a secondary stage that sets in later but with a significantly larger growth rate. In particular, for not too high momenta the primary growth abruptly changes and a secondary, larger growth rate takes over as exemplified for $p = 0.8 \, \epsilon^{1/4}$ in Fig.~\ref{A-vs-t}. The observation of primary and secondary growth rates closely resembles findings in the context of early universe preheating dynamics in self-interacting scalar inflaton models following parametric resonance or spinodal instabilities~\cite{Berges:2002cz}. Similar to the analysis for inflaton dynamics, we will demonstrate below in section~\ref{diagram-section} that the secondaries arise from fluctuation effects induced by the growth in the lower momentum modes by taking into account resummed loop diagrams beyond the hard-loop approximation. This growth saturates when all loop diagrams are of order one corresponding to a highly non-linear evolution.\footnote{This corresponds to $\langle A A \rangle \sim 1/g^2$ in a description where the coupling would not be scaled out as described in Sec.~\ref{sec:classical-statistical}. See also the discussion in Ref.~\cite{Berges:2004yj,Berges:2002cz}.} For the calculation for Fig.~\ref{A-vs-t}, which employs $\Delta/\epsilon^{1/4} = 1 $, saturation occurs around $t \sim 160 \, \epsmf$. 

From Fig.~\ref{A-vs-t} one also concludes that the same phenomena could be observed if the simulation started at a later time than $t_0$. Choosing the zero point of the time axis just changes the details of the initial conditions. For instance, even at times $\Delta t = 50 \, \epsilon^{-1/4}$ or $100 \, \epsilon^{-1/4}$ the anisotropy is still substantial. As an indicator one may consider
the bulk anisotropy parameter $\xi(t)$ defined as 
\begin{equation}\label{def-xi-t}
\xi(t) \equiv \log_{10} \left\{ \, \frac{ \frac{1}{2} \, \sum_{\vec{p}} \, ( p_x^2 + p_y^2 ) \, \Bigl( \, \sum_{j=1}^{3} \, \sum_{a=1}^3 \, | \, A_j^a(t, \vec{p}) \, |^2  \, \Bigr) }{ \sum_{\vec{q}} \, q_z^2 \, \Bigl( \, \sum_{k=1}^{3} \, \sum_{b=1}^3 \, | \, A_k^b(t, \vec{q}) \, |^2 \Bigr) } \right\} \, .
\end{equation}
The points when $\xi(t)$ takes particular values are indicated in Fig.~\ref{A-vs-t} by dashed vertical lines. One can deduce from this that the configurations up to approximately $\Delta t \sim 120 \epsmf$ would equally well qualify as initial conditions with a very high degree of anisotropy. In order to extract characteristic time scales for isotropization we, therefore, focus on the inverse growth rates.

\begin{figure}[t!]
\begin{center}
\epsfig{file=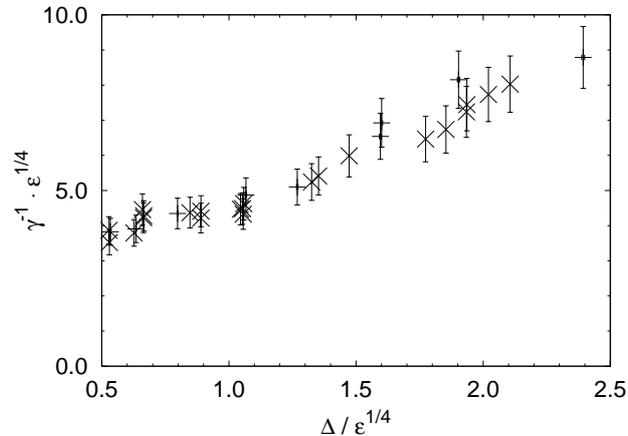, width = \mywidth }
\caption{Time scales associated with the maximum primary growth rate as a function of the transverse width $\Delta$ in units of $\epsmf$.}
\label{fig-growth-rates-vs-dimless-ratio}
\end{center}
\end{figure}
Fig.~\ref{fig-growth-rates} displays the momentum dependence of the growth rates for $| A(t, \vec{p}) |^2$ obtained from a fit to an exponential. According to the above discussion this is done separately for the primary and secondary growth rates. One observes that the secondaries occurring at higher momenta exhibit growth rates which are typically larger than the highest primary rate by a factor of approximately three to four. The dependence of the results on the initial width $\Delta$ in the transverse plane can be inferred from Fig.~\ref{fig-growth-rates-vs-dimless-ratio}, where the inverse of the maximum primary growth rate as a function of $\Delta / \epsf$
is plotted. 
 
\begin{figure}[t!]
\begin{center}
\epsfig{file=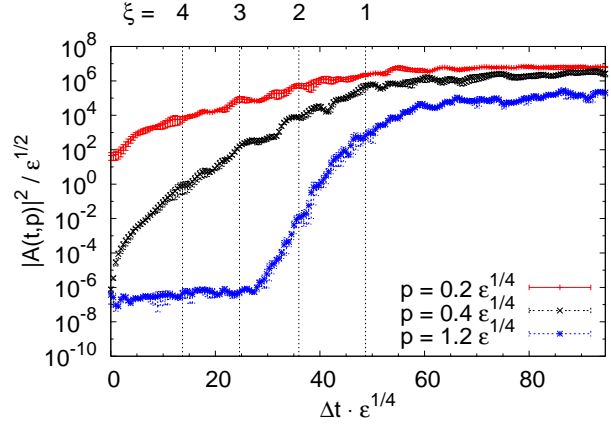, width= \mywidth }
\caption{Time evolution of $|\, A(t, \vec{p}) \, |^2$ from initial conditions with a slightly broadened distribution in the longitudinal direction with $\Delta/\Delta_z = 25$. The curves correspond to the momenta in the legend from top to bottom in the same order.}
\label{old-initial-conditions-gauge-field}
\end{center}
\end{figure}
\begin{figure}[t!]
\begin{center}
\epsfig{file=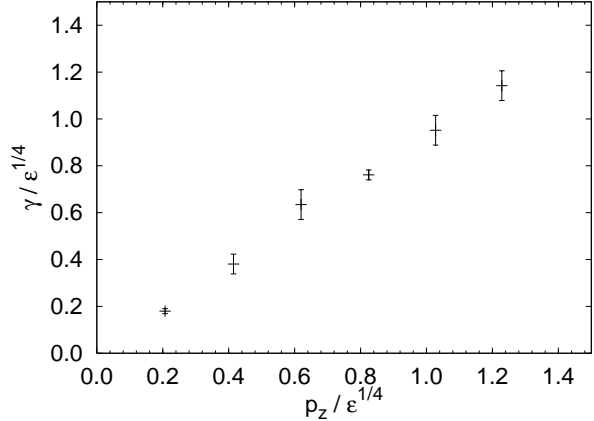, width = \mywidth}
\caption{Growth rates of $| \, A(t, \vec{p}) \, |^2 $ as a function of $p_z$ corresponding to Fig.~\ref{old-initial-conditions-gauge-field}.}
\label{old-initial-conditions-growth-rate}
\end{center}
\end{figure}

 
For the conversion to physical units we will consider two different values for the energy density of about 1 GeV/fm$^3$ and 30 GeV/fm$^3$, equivalent to approximately ($300$ MeV)$^4$ $\simeq$ ($1.5/$fm)$^4$ and (700 MeV)$^4$ $\simeq$ ($3.5/$fm)$^4$. For RHIC experiments the former estimate is expected to be too low, while the latter can be still appropriate~\cite{Gyulassy:2004zy}. According to Fig.~\ref{fig-growth-rates-vs-dimless-ratio}, the inverse of the maximum primary growth rate for $|\, A(t, \vec{p}) \, |^2$ is $ \gamma_{\text{max. pr.}}^{-1} \simeq 4 \, \epsmf$. In physical units this corresponds to 
\begin{eqnarray}\label{optimistic-inverse-primary-gr}
 \gamma_{\text{max.\ pr.}}^{-1}  &\simeq& 1.1 \, \text{fm}/\text{c} \hspace{1cm} \text{($\epsilon$ = 30 GeV/fm$^3$)}\, , \\ 
 \gamma_{\text{max.\ pr.}}^{-1}  &\simeq& \, 2.7 \text{fm}/\text{c} \hspace{1cm} \text{($\epsilon$ = 1 GeV/fm$^3$).}
\label{pessimistic-inverse-primary-gr}
\end{eqnarray}
In order to achieve growth by a few orders of magnitude in the low-momentum occupation numbers a time of several $\gamma_{\text{max. pr.}}^{-1}$ has to elapse. One may then ask what energy density would be required to get characteristic inverse growth rates of about $0.1$fm/c and we obtain:
\begin{eqnarray}\label{bound}
 \gamma_{\text{max. pr.}}^{-1}  &\simeq& 0.1 \, \text{fm}/\text{c} \hspace{1cm} \text{($\epsilon$ = 300 TeV/fm$^3$}\, \text{ !)} 
\end{eqnarray}
The situation changes somewhat if one considers the characteristic time scales associated with the secondary growth rates. For these we find that $\gamma_{\text{max. sec.}}^{-1} \simeq 1.3 \, \epsmf$, which in physical units corresponds to 
\begin{eqnarray}\label{optimistic-inverse-secondary-gr}
 \gamma_{\text{max.\ sec.}}^{-1}  &\simeq& 0.4 \, \text{fm}/\text{c} \hspace{1cm} \text{($\epsilon$ = 30 GeV/fm$^3$)}\, , \\ 
 \gamma_{\text{max.\ sec.}}^{-1}  &\simeq&  0.8 \, \text{fm}/\text{c} \hspace{1cm} \text{($\epsilon$ = 1 GeV/fm$^3$).}
\label{pessimistic-inverse-secondary-gr}
\end{eqnarray}
At first sight, these seem to be easier to reconcile with the envisioned fast isotropi\-zation. However, even though secondaries can reach much higher growth rates their appearance can be delayed. According to the results in Fig.~\ref{A-vs-t} the secondaries only affect the isotropization process at early times if the initial conditions happen to be similar to the gauge field configuration at $\Delta t \simeq 100 - 120 \epsmf$ because otherwise they would set in too late. So far, we have initialized our runs using a $\delta(p_z)$-like distribution at $t=t_0$. Other initial conditions can provide for a more pronounced role of secondaries. An example is given in Fig.~\ref{old-initial-conditions-gauge-field}, which shows the time evolution of $|\, A(t, \vec{p}) \, |^2$ starting from the same Gaussian initial condition as given in Eq.~(\ref{InitCond-1}), however, for a slightly broadened distribution in the longitudinal direction with $\Delta/\Delta_z = 25$. Indeed, one sees that from the very beginning growth rates comparable to the secondary instabilities observed in Fig.~\ref{A-vs-t} occur. In Fig.~\ref{old-initial-conditions-growth-rate} these rates are given as a function of $p_z$. Comparison with Fig.~\ref{fig-growth-rates} for given momenta shows that the growth is even faster by a factor of approximately two. This would lead to somewhat shorter time scales than presented in Eqs.~(\ref{optimistic-inverse-primary-gr})-(\ref{pessimistic-inverse-secondary-gr}). 
The analysis of (primary) growth rates using hard loop approximations leads to the expectation that the rates decrease with decreasing $\Delta/\Delta_z$, i.e.~the less anisotropic the initial conditions are. The above results show that (secondary) growth rates can, however, increase with decreasing $\Delta/\Delta_z$. Their understanding requires to go beyond hard loop approximations, which is done in the following.

\subsection{The impact of fluctuations: Diagrammatic analysis}\label{diagram-section}

We have seen above that low momentum modes of $| \, A(t, \vec{p}) \, |^2$ show an exponential increase starting at the very beginning of the simulation, while for sufficiently large initial anisotropy higher momentum modes exhibit exponential growth at a secondary stage with rates several times larger than primary ones. Following the discussion of qualitatively similar findings in Refs.~\cite{Berges:2004yj,Berges:2002cz} for self-interacting scalar preheating dynamics, we analyze the appearance of secondaries in terms of resummed self-energy diagrams. These diagrams appear as source terms in the Schwinger-Dyson equations for propagators. They naturally arise using two-particle irreducible (2PI) effective action techniques for the description of nonequilibrium dynamics of correlators such as the equal-time two-point function~\cite{Berges:2004yj} 
\begin{equation}\label{def-correlation-function}
F_{\mu \nu}^{ab}(t, \vec{x} - \vec{y}) = \langle \, A_{\mu}^a(t, \vec{x}) A_{\nu}^b(t, \vec{y}) \, \rangle \, .
\end{equation}
Since we work in temporal axial gauge, i.e. $ A_0^a(x) \equiv 0 $, it follows that $F_{00}^{ab} = F_{0\mu}^{ab} = F_{\mu 0}^{ab} = 0 $. 

\begin{figure}[t!]
\begin{center}
\subfigure[]{\epsfig{file=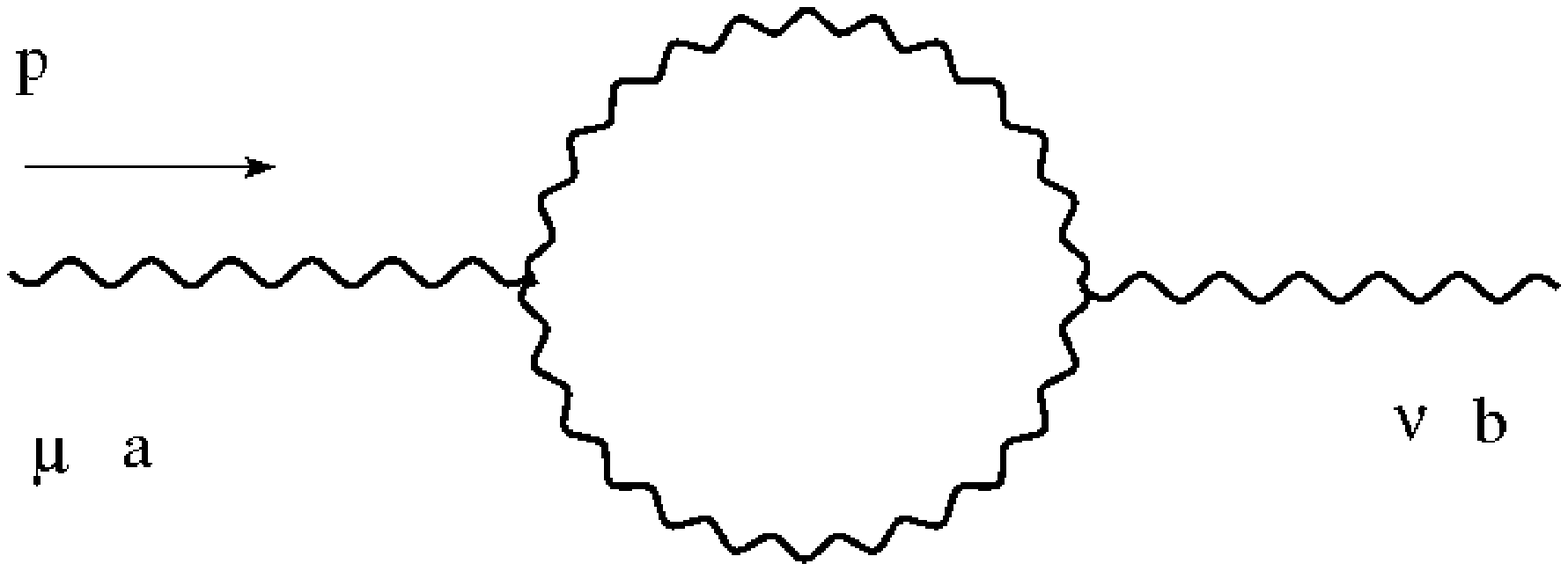, width=6.8cm}}
\subfigure[]{\epsfig{file=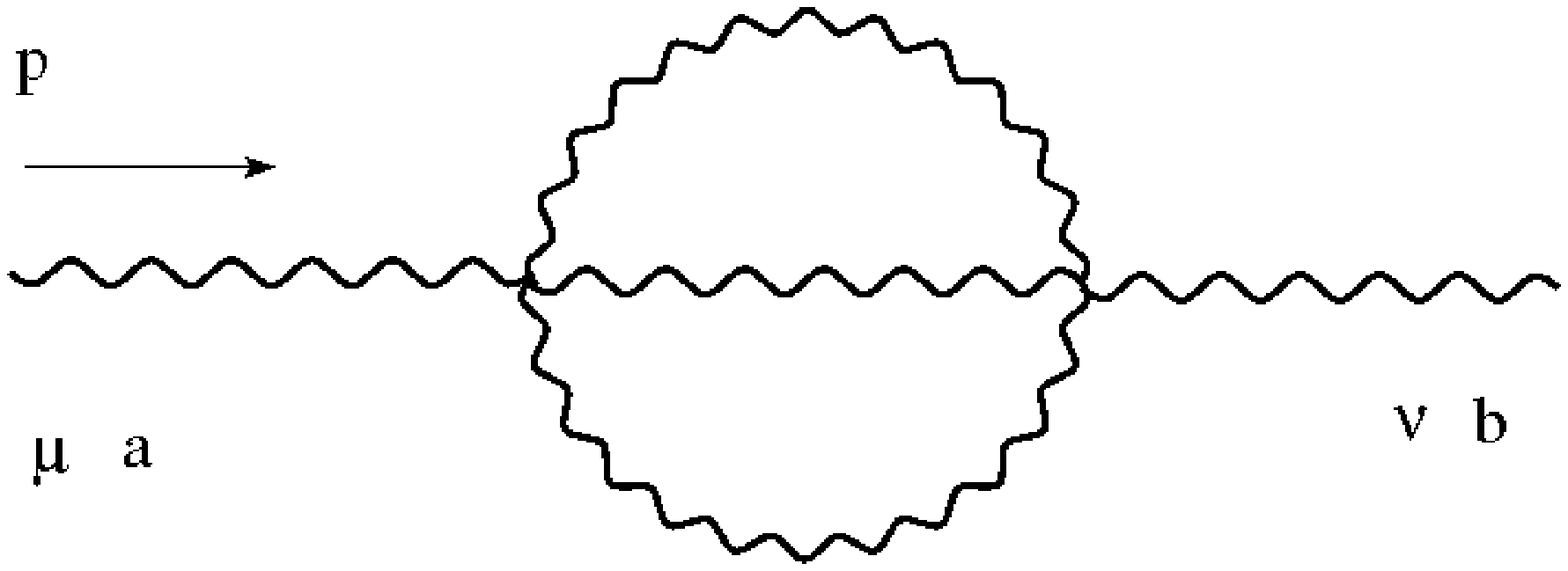, width=6.8cm}}
\caption{Loop diagrams (a) $\Pi^{\mu \nu \, (1)}_{ab}(t, \vec{p})$ given in Eq.~\eqref{def-one-loop} and (b) $\Pi^{\mu \nu \, (2)}_{ab}(t, \vec{p})$ given in Eq.~\eqref{def-setting-sun}. The propagator lines correspond to the exact classical-statistical propagator, while the vertices are the perturbative ones.}
\label{sketch_diagrams}
\end{center}
\end{figure}

Low-order diagrams in a loop expansion of the self-energy contributing to the correlator (\ref{def-correlation-function}) are displayed in Fig.~\ref{sketch_diagrams}.
We emphasize that the diagrams are computed using the exact classical-statistical propagator obtained from simulations and perturbative vertices~\cite{Berges:2004yj, PRD2004}. The perturbative three- and four-gluon vertices are given as  
\begin{equation}\label{def-V3}
V^{\mu \nu \rho}_{3 \, a b c}(\vec{k}, \vec{p}, \vec{q}) =  g \epsilon_{abc} \, \bigl\{ g^{\mu \nu} ( \vec{k} -\vec{p})^{\rho}  +g^{\nu \rho}(\vec{p} -\vec{q})^{\mu} + g^{\rho \mu}( \vec{q} - \vec{k} )^{\nu} \bigr\}
\end{equation}
and 
\begin{equation}\label{def-V4}
\begin{split}
 V^{\mu \nu \rho \sigma}_{4 \, a b c d} = - i g^2 \bigl\{ & \epsilon_{abf} \epsilon_{cdf}( g^{\mu \rho}g^{\nu \sigma} - g^{\mu \sigma} g^{\nu \rho}  ) \\
& + \epsilon_{acf} \epsilon_{bdf}( g^{\mu \nu }g^{\rho \sigma } - g^{\mu \sigma} g^{\nu \rho} ) + \epsilon_{adf} \epsilon_{bcf}( g^{\mu \nu }g^{\rho \sigma} - g^{\mu \rho} g^{\nu \sigma } ) \bigr\}  \; .
\end{split}
\end{equation}
In the above vertices the Lorentz indices are restricted to run over spatial directions only (i.\ e. $\mu = \{1, 2, 3 \}$) since temporal components only appear together with vanishing $F_{0 \mu}^{a b}$ and $F_{00}^{ab}$. The corresponding contribution of the one loop diagram Fig.~\ref{sketch_diagrams}(a) reads   
\begin{equation}\label{def-one-loop}
\begin{split}
\Pi^{\nu \sigma \, (1) }_{b f}(t, \vec{p}) = \frac{1}{2} \int \frac{d^3q}{(2 \pi)^3}\, \Bigl\{\, & V^{\mu \nu \rho}_{3 \, abc}( - \vec{p} - \vec{q}, \vec{p}, \vec{q}) V^{\kappa \sigma \lambda}_{3 \, dfe}(\vec{p} + \vec{q}, -\vec{p}, - \vec{q}) \\ 
&  \times \, F^{da}_{\kappa \mu}(t, \vec{p} + \vec{q}) F^{ec}_{\lambda\rho}(t, -\vec{q}) \, \Bigr\}  \; .
\end{split}
\end{equation}
The two-loop diagram Fig.~\ref{sketch_diagrams}(b) is evaluated in coordinate space according to 
\begin{equation}\label{def-setting-sun}
\Pi^{\mu \nu \, (2)}_{m n}(t, \vec{x} - \vec{y}) = \frac{1}{6} \, V^{\mu \alpha \beta \gamma }_{4 \, mabc} V^{\nu \kappa \lambda \rho }_{4 \, nklr} F^{\kappa \alpha}_{k a}(t, \vec{x} - \vec{y}) F^{\lambda \beta}_{l b}(t, \vec{x} - \vec{y}) F^{\rho \gamma}_{r c}(t, \vec{x} - \vec{y}) 
\end{equation}
and then Fourier transformed.

\begin{figure}[t!]
\begin{center}
\epsfig{file=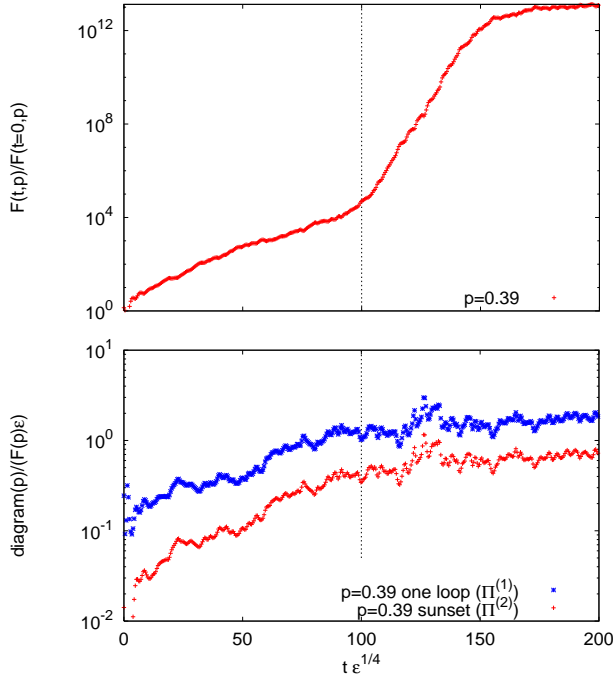, width=\mywidth}
\caption{The upper graph displays the correlation function $F(t, \vec{p})$ defined in Eq.~(\ref{def-correlation-function}) as a function of time for given momentum $p_z/\epsilon^{1/4} = 1$. The lower graph shows for the corresponding momentum the relative contributions of the diagrams $ \Pi^{(1)}(t, \vec{p})/ F(t, \vec{p}) $ and $\Pi^{(2)}(t, \vec{p})/ F(t, \vec{p})$ to the correlation function. The upper curve gives the one-loop contribution, the lower one is the setting-sun contribution.}
\label{diagrams}
\end{center}
\end{figure}
Fig.~\ref{diagrams} shows the results of such a calculation where all external color indices have been set to one and Lorentz indices were chosen in the transverse plane, i.e.\ perpendicular to the axis of anisotropy. The upper graph displays $F^{11}_{11}(t, \vec{p})$ for a given momentum as a function of time, similar to Fig.~\ref{A-vs-t}. The momentum $p_z/\epsilon^{1/4} = 1$ is chosen such that the mode exhibits a primary growth rate initially which abruptly changes at a later time of about $\Delta t \epsilon^{1/4} = 100$ to a secondary rate. The lower graph of Fig.~\ref{diagrams} gives the absolute values of the loop diagrams Eqs.~\eqref{def-setting-sun} and \eqref{def-one-loop} normalized to the correlation function Eq.~\eqref{def-correlation-function}. The diagrams are evaluated for the same momentum as the gauge fields in the upper graph. One observes that initially these loop corrections are small compared to the size of the correlator (\ref{def-correlation-function}) in units of $\epsilon$. Accordingly, their contribution to the time evolution of the correlator may be neglected and primary growth rates appear as a consequence of Weibel instabilities as discussed in great detail in the literature~\cite{Mrowczynski:1988dz}. However, due to these instabilities loop corrections grow exponentially and faster than the correlator itself. The faster growth is seen to stop at about $\Delta t = 100 \, \epsilon^{-1/4}$. This is precisely the time where the secondary growth rate for the correlator sets in. The faster growth of the loop corrections originates from the fact that the exponentially growing correlation modes $F(t,\vec{p})$ appear multiple times as propagators in the loop diagrams such as shown in Fig.~\ref{sketch_diagrams}. For the case of the scalar preheating dynamics analyzed in Refs.~\cite{Berges:2004yj,Berges:2002cz} it was shown that secondary growth rates indeed equal multiples of primary ones. This was possible because primary growth was sharply peaked around certain momenta in the scalar models at weak coupling, and the loop integrals could be estimated in a saddle point approximation around the dominating loop-momentum. For the SU(2) gauge theory considered here primaries do not appear in a sharply peaked momentum range as can be observed from Fig.~\ref{fig-growth-rates}. Accordingly, we do not find that secondary growth rates correspond to simple multiples of primary ones.  

In section \ref{sec:growth-rates} we have seen that for large anisotropies, corresponding to $\delta(p_z)$-like initial distributions (\ref{InitCond-1}), sizeable secondaries only appear at later times and not from the beginning of the simulation. This can be understood from the fact that for these initial conditions the loop corrections as in Fig.~\ref{sketch_diagrams} are negligible at initial time for longitudinal momenta. In contrast, broadening of the initial distribution in the longitudinal direction can lead to sizeable initial loop corrections. Accordingly, we find for such initial conditions larger growth rates already at earliest times as demonstrated in Figs.~\ref{old-initial-conditions-gauge-field} and \ref{old-initial-conditions-growth-rate}. However, from comparing Fig.~\ref{old-initial-conditions-gauge-field} with Fig.~\ref{A-vs-t} one observes that the exponential behavior is less developed the less anisotropic the initial conditions. As a consequence, we find that characteristic results for the isotropization of observables such as pressure are rather insensitive to the details of the considered initial conditions which is discussed in the following section.

\section{Bottom-up isotropization}\label{sec:bottom-up-isotropization}

\begin{figure}[!t]
\begin{center}
\epsfig{file=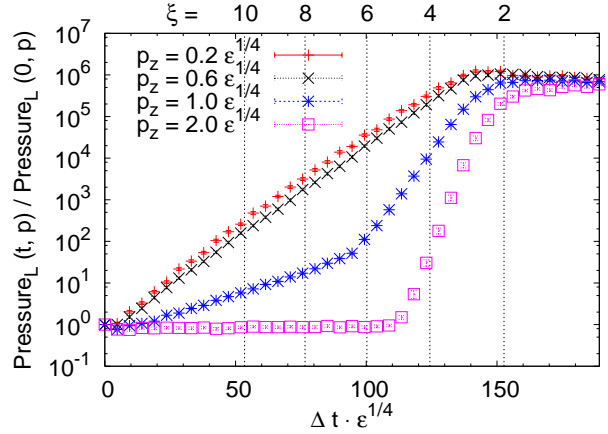 , width = \mywidth }
\caption{Time evolution of longitudinal "pressure", i.e.\ the spatially Fourier transformed longitudinal component of the energy-momentum tensor $T_{33}(t, \vec{p})$ normalized to initial values for various momenta $p_z$.}
\label{fig:pressure}
\end{center}
\end{figure}
\begin{figure}[!t]
\begin{center}
\epsfig{file=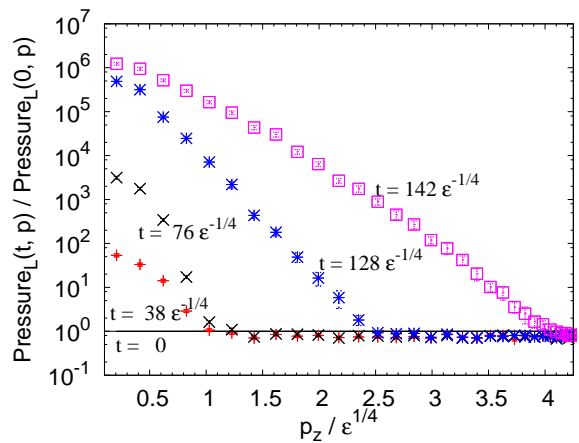, width = \mywidth } 
\caption{Snapshots of the spectrum of $T_{33}(t, \vec{p}) / T_{33}(t=0, \vec{p})$ at various times.}
\label{pz-spectra}
\end{center}
\end{figure}

We compute the time evolution of the spatial components of the energy-momentum tensor with initial conditions as employed for Figs.~\ref{A-vs-t} - \ref{fig-growth-rates-vs-dimless-ratio}. Its diagonal components yield the pressure in thermal equilibrium and we will refer to $T_{11}$ and $T_{22}$ as the time-dependent transverse pressure. Accordingly, $T_{33}$ will be called the longitudinal pressure. The diagonal components of $T_{ij}$ can be computed according to 
\begin{equation}
\begin{split}
T_{33}(x) =  - \,  \frac{1}{2 a_s^4 g^2} \, \Bigl\{   \gamma^2 \Bigl( \bigl( 1 - \frac{1}{2} \, \trace \, U_{x, 01} \, \bigr) + \bigl( 1 - \frac{1}{2} \, \trace \, U_{x, 02} \, \bigr) - \bigl( 1 - \frac{1}{2} \, \trace \, U_{x, 03} \, \bigr) \Bigr)  \\
  +  \, \bigl(  1 - \frac{1}{2} \, \trace \, U_{x, 23} \, \bigr) + \bigl(  1 - \frac{1}{2} \, \trace \, U_{x, 31} \, \bigr) - \bigl(  1 - \frac{1}{2} \, \trace \, U_{x, 12} \, \bigr)  \Bigr\}  \, ,
\end{split}
\end{equation}
and $T_{11}$ and $T_{22}$ can be obtained by permuting the indices respectively. We have verified that non-diagonal components of the energy-momentum tensor are negligible at all times for the class of initial conditions considered here. We then Fourier transform w.r.t.\ the spatial components and consider the momentum modes of the longitudinal and transverse pressure as a function of time. Fig.~\ref{fig:pressure} shows the longitudinal pressure modes for four different momenta in the z-direction, normalized to the respective initial values. One observes that the longitudinal pressure - which is a gauge invariant quantity - exhibits instabilities in a similar way as does the gauge field. The same qualitative picture regarding time scales, primary and secondary instabilities emerges from Fig.~\ref{fig:pressure} as from Fig.~\ref{A-vs-t}. Quantitatively, a closer look reveals that the growth rates of modes of $T_{33}$ are smaller by a factor of about two than the corresponding modes in $| \, A(\vec{p}) \, |^2$ with the same wave number. This can be understood from the fact that the longitudinal pressure contains contributions $\sum_{\vec{q}} \, A(t, \vec{p}) A(t, \vec{q} - \vec{p})$, which are dominated before saturation by momenta where $\vec{q} - \vec{p}$ lies in the transverse plane. Since the transverse modes do not exhibit instabilities, there is only one exponentially growing factor appearing in each of the dominant contributions.
Fig.~\ref{pz-spectra} shows the same quantity as in Fig.~\ref{fig:pressure}, however, as a function of longitudinal momentum. Plotted are snapshots of the spectrum at fixed times versus the absolute value of the momentum. One observes a band of low-momentum modes that are unstable at initial times. At the time when the secondary instabilities set in also a broad regime of higher modes suddenly starts to grow, similar to what has been described in Ref.~\cite{Arnold:2005ef} and in the second paper of Ref.~\cite{WongYangMills}.   

\begin{figure}[!t] 	
\begin{center}
\epsfig{file=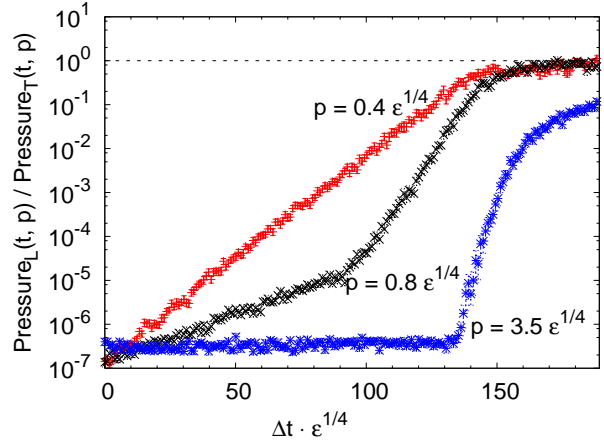 , width = \mywidth}
\caption{The ratio of modes of longitudinal pressure and transversal pressure (\ref{ratio-T-comps}) as a function of time for different momenta. One observes that pressure isotropizes "bottom-up" and complete isotropization does not occur on short time scales.}
\label{bottom-up-isotropization}
\end{center}
\end{figure}
The momentum scale of the Fourier transformed spatial components of the energy-momentum tensor may be viewed as a characteristic scale of a coarse grained pressure. One can then ask the question about a characteristic coarse graining scale up to which fast isotropization may occur. Fig.~\ref{bottom-up-isotropization} shows the ratio of longitudinal and transverse modes of pressure, i.e.\
\begin{equation}\label{ratio-T-comps}
\Bigl| \frac{ T_{33}(t, \vec{p}_{||}) }{T_{\perp}(t, \vec{p}_{\perp}) }  \Bigr|_{| \, \vec{p}_{||} \, | = | \, \vec{p}_{\perp} \, | = p } \, ,
\end{equation}
for three different momenta as a function of time. One observes that for low-momentum modes the ratio turns to one rather quickly. High momentum modes only become isotropic at far later times.  Complete isotropization does not occur on short time scales. The system becomes only effectively isotropic when considered on sufficiently large characteristic length scales. This may be used to define an isotropic coarse grained pressure by averaging over a volume with a characteristic length scale given by the inverse momentum. For instance, requiring that $P_L(\vec{p}) / P_T(\vec{p})\gtrsim 0.6 $ at the time when the instabilities saturate ($\Delta t = 160 \epsmf$ in Fig.~\ref{A-vs-t}) we find isotropization up to momenta $p_z \lesssim 1.4 \epsf $, which corresponds to 
\begin{eqnarray}
 p_z &\lesssim& 1 \text{GeV} \hspace{1.5cm} \text{($\epsilon$ = 30 GeV/fm$^3$)}\, , \\ 
 p_z &\lesssim& 400 \text{MeV} \hspace{1cm} \text{($\epsilon$ = 1 GeV/fm$^3$).}
\label{eq:cgpressure}
\end{eqnarray}
Hydrodynamic simulations indeed assume locally isotropic pressure only on sufficiently large characteristic length scales~\cite{Heinz:2004pj}. However, even for $\epsilon$ = 30 GeV/fm$^3$ the maximum coarse graining momentum we find is somewhat small. These results are obtained for the $\delta(p_z)$-like initial conditions as explained above, and in view of the results of section~\ref{Results} one might expect that broadening the initial distribution in the longitudinal direction improves the situation and leads to earlier isotropization of a broader momentum range. However, we find that the maximum mode that has a ratio of $P_L(\vec{p}) / P_T(\vec{p})\gtrsim 0.6 $ at saturation has a momentum of about $p_z \simeq 1.5 \epsilon^{1/4}$ for the less anisotropic initial conditions employed for Figs.~\ref{old-initial-conditions-gauge-field} and \ref{old-initial-conditions-growth-rate}. Despite the fact that growth rates are twice as large in this case, the exponential behavior is less developed the less anisotropic the initial conditions. We conclude that the estimates (\ref{eq:cgpressure}) are rather insensitive to the details of the considered initial conditions. 

\section{Wilson loops}\label{wilson-loop-section}

Wilson loops as traces of parallel transporters around rectangular loops provide an important means to describe confinement properties of non-Abelian gauge theories. In vacuum large Wilson loops with side length $R$ and $\tilde{R}$ obey the area law 
\begin{equation}
 W(R,\tilde{R}) \,\sim\, e^{- \kappa R \tilde{R}} \, 
\label{wloop}
\end{equation}
for $R,\tilde{R} \to \infty$. Here $\kappa$ denotes the string tension and a non-vanishing value can be related to static quark confinement for loops including a time separation on a Euclidean space-time lattice. For the latter it is known that spatial Wilson loops, where $R$ and $\tilde{R}$ both denote spatial distances, give the same value for $\kappa$ as the temporal ones and that the "spatial string tension" depends rather weakly on the temperature \cite{Manousakis:1986jh}.   
\begin{figure}[t!]
\begin{center}
\epsfig{file=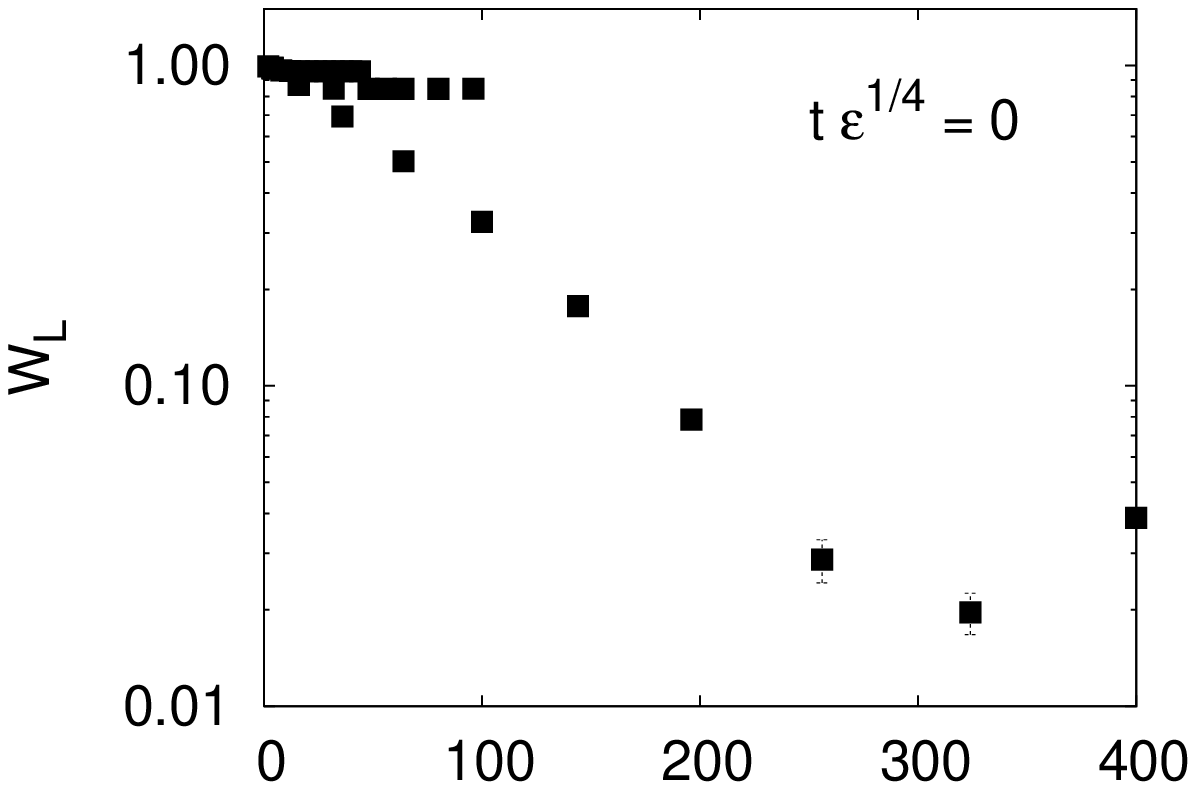, width=4.5cm}
\epsfig{file=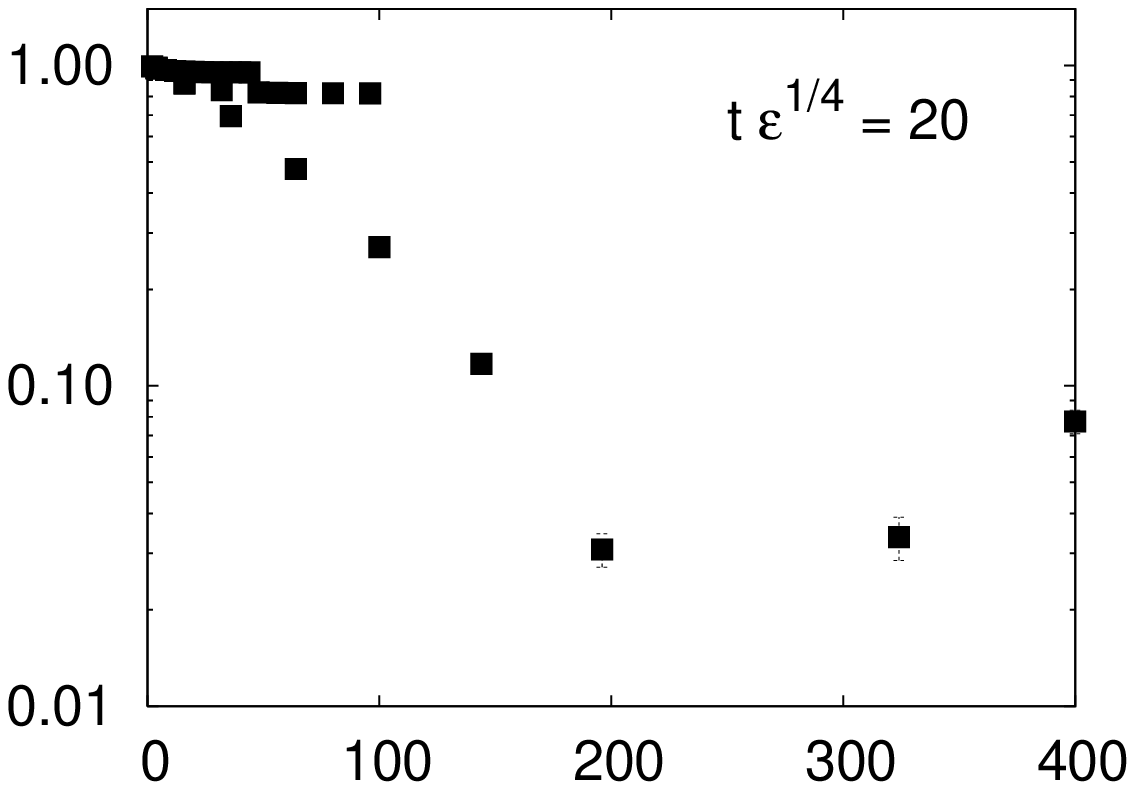, width=4.5cm}
\epsfig{file=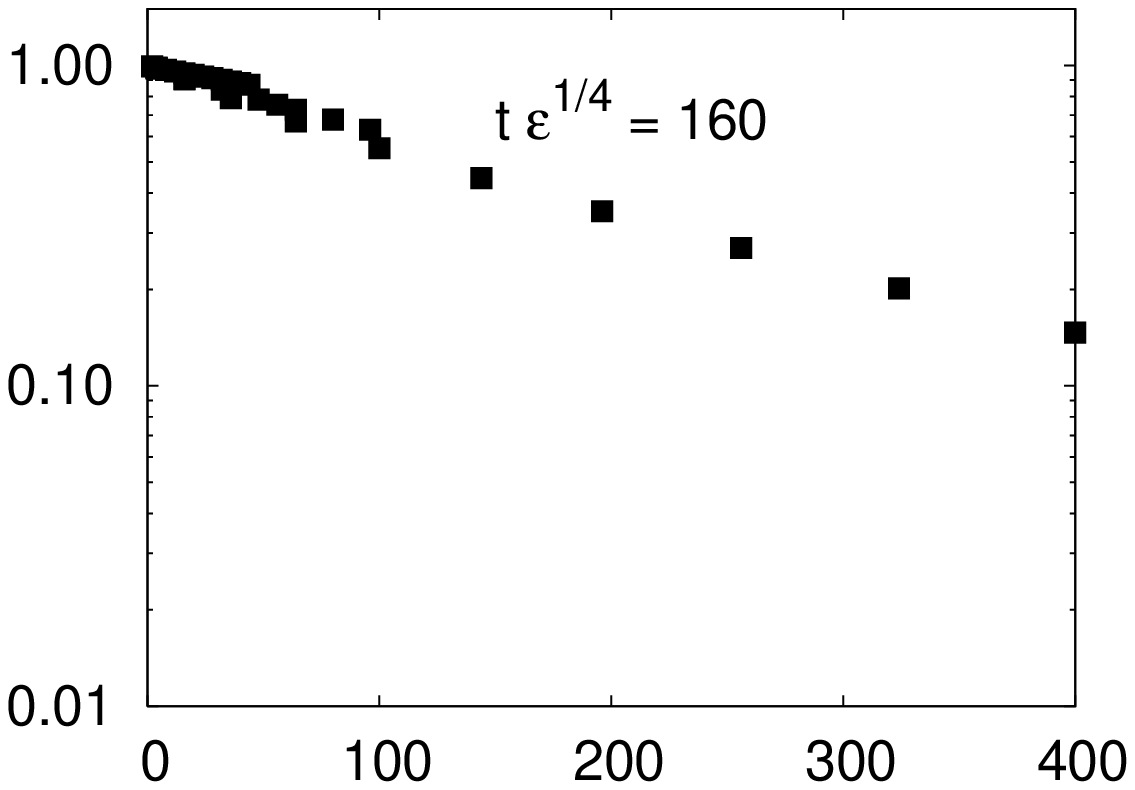, width=4.5cm}
\epsfig{file=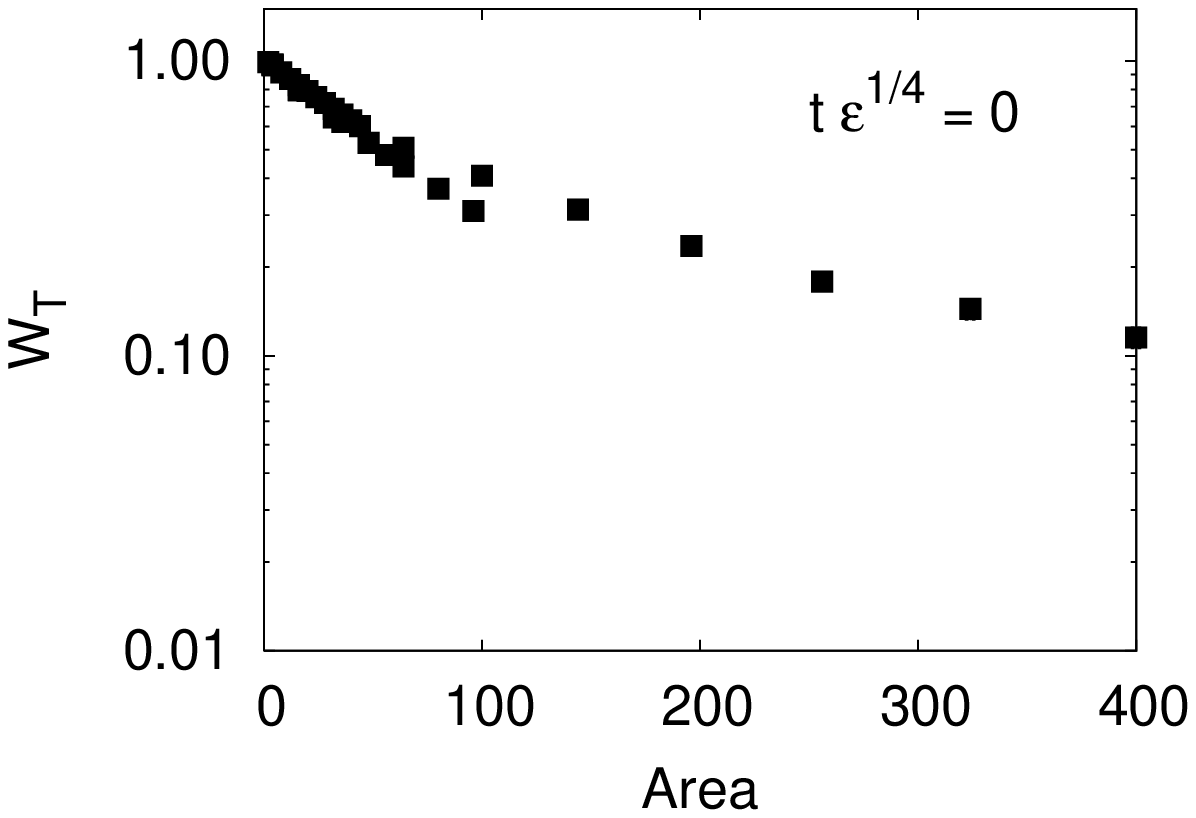, width=4.5cm}
\epsfig{file=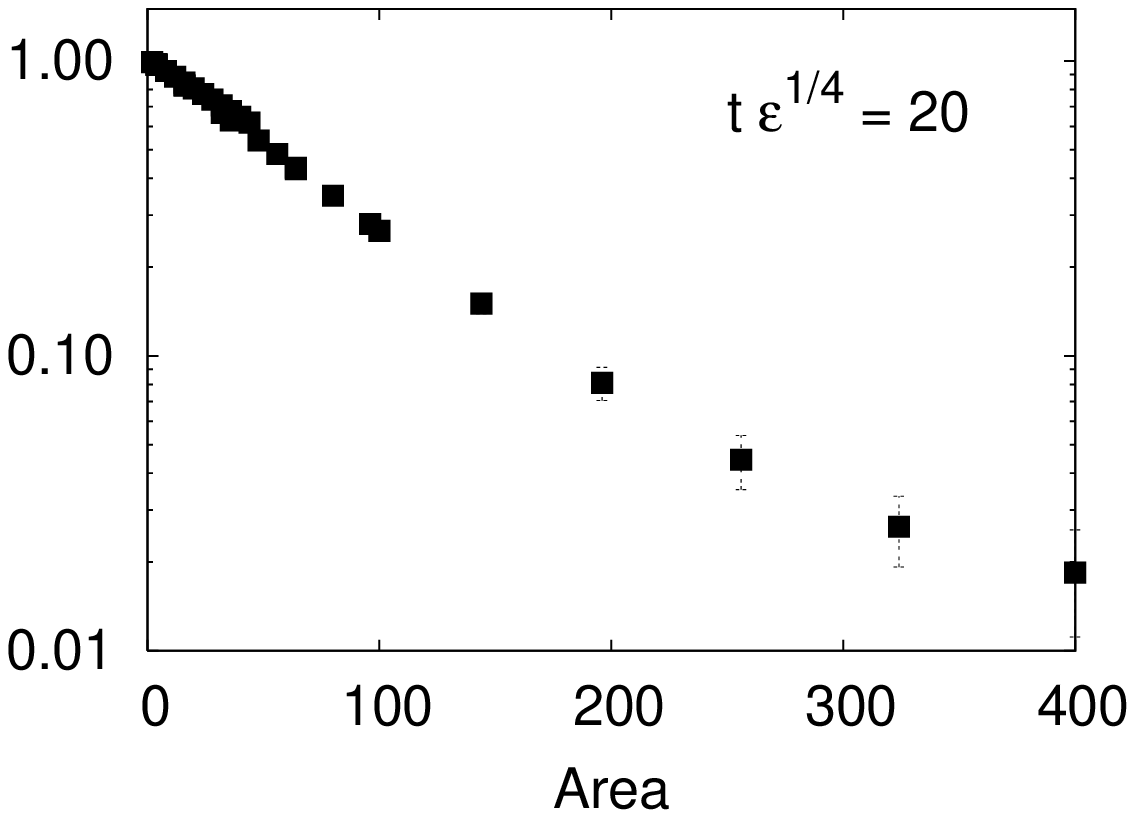, width=4.5cm}
\epsfig{file=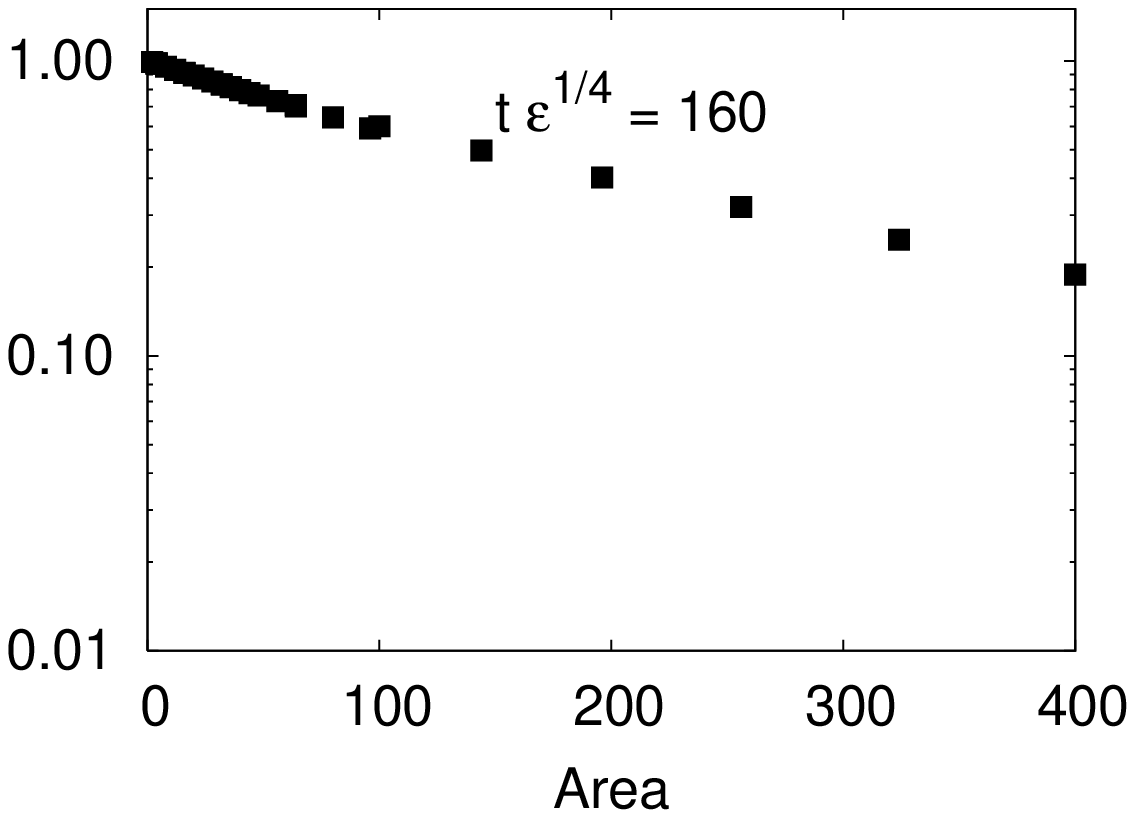, width=4.5cm}
\caption{Values of spatial Wilson loops in the transverse plane ($W_T$) and Wilson loops including the longitudinal direction ($W_L$) as a function of the loop area in lattice units. Shown are snapshots at increasing time going to the right. Initially, $W_L$ exhibits no area law, while the transverse Wilson loop is characterized by a non-vanishing spatial string tension $\kappa$ as explained in the text. One observes $W_L/W_T \simeq 1$ at the time of saturation of the exponential growth of gauge field fluctuations and the anisotropy is removed. }
\label{fig-stringstory}
\end{center}
\end{figure}

On a real-time lattice the nonequilibrium time evolution of spatial Wilson loops can be measured and the corresponding spatial string tension can be extracted according to Eq.~(\ref{wloop}). We measure the time evolution of large Wilson loops in the transverse plane, i.e.\ with $R$ and $\tilde{R}$ measuring distances in the $xy$-plane, and loops where one side length measures distance in the $z$-direction. We will call the former transverse Wilson loops, $W_T$, and the latter longitudinal Wilson loops, $W_L$, accordingly. We emphasize that since we are considering classical-statistical time evolution we do not take into account quantum fluctuations. This can only be a good approximation if classical-statistical fluctuations dominate. While this is less problematic for bulk quantities such as pressure, which is not dominated by low-momentum fluctuations, this is in general not justified for a quantity such as the string tension that is sensitive to long-ranged fluctuations related to confinement. Accordingly, the longitudinal Wilson loops do not show an area law at initial time for the class of initial conditions considered here because of the low occupation numbers in the longitudinal direction. 

It is remarkable, however, that the transverse Wilson loops exhibit an area law as a consequence of large classical-statistical fluctuations in the presence of the high initial transverse occupation numbers. This is demonstrated in Fig.~\ref{fig-stringstory}, where the values for $W_L$ and $W_T$ are given at various times as a function of the area $R \times \tilde{R}$ in lattice units. We evaluate a variety of rectangular loops with one side length $R=2 , 4$ or $6$, while the other side length is varied in the range $\tilde{R} \in \{2,4,...,20\}$ and square-shaped Wilson loops with $R = \tilde{R}$. One observes that initially $W_L$ exhibits no area law. In contrast, the values of the transverse Wilson loops follow an area law to very good accuracy. The initial conditions are as described in section~\ref{sec:classical-statistical} for a $\delta(p_z)$-like initial distribution and we employ a lattice energy density $\hat{\epsilon} = 0.05$ according to Eq.~(\ref{eps-epshat}) and a transverse width $\hat{\Delta} = 0.5$ in lattice units on a $64^3$ lattice. The figures to the right show the same quantities at later times. The longitudinal Wilson loop is seen to develop an area law and at the time $t \simeq 160 \epsmf $, after the exponential growth of gauge field fluctuations saturates, all Wilson loops follow the same area law and the anisotropy is removed with $W_L/W_T \simeq 1$. We conclude that the isotropization of the spatial Wilson loops occurs on a similar characteristic time scale as the low-momentum coarse-grained pressure.

\begin{figure}[t!]
\begin{center}
\epsfig{file=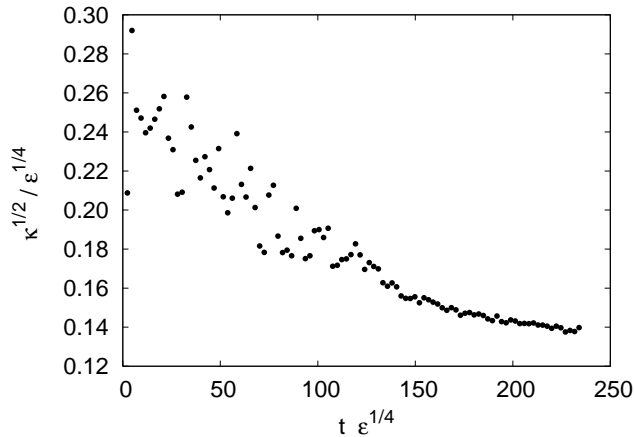, width = \mywidth}
\caption{The time dependence of the spatial string tension from Wilson loops
in the transverse plane.}
\label{fig-stringtension}
\end{center}
\end{figure}

Though the string tension extracted from the transversal Wilson loops is non-vanishing from the beginning, it changes with time. In Fig.\ref{fig-stringtension} the corresponding value of $\kappa^{1/2}$ is plotted as a function of time in units of the energy density $\epsilon$. For the considered initial conditions one observes an initial decrease with time, which tends to a non-zero value $ \kappa^{1/2} / \epsf \simeq 0.14 $ at late times. We verify the expectation that quantities such as pressure are not sensitive to this scale. As described in sections \ref{Results} and \ref{wilson-loop-section} the pressure as well as the growth rates only depend on $\Delta / \epsf $ for a given time or momentum.
      
The result of a non-vanishing scale as defined by Eq.~(\ref{wloop}) for spatial Wilson loops may seem surprising in a classical-statistical theory at first sight. However, our nonequilibrium results relate well to known behavior in thermal equilibrium at high temperature. The low-momentum behavior of a $3+1$-dimen\-sional bosonic quantum theory at sufficiently high-temperature is described in terms of a $3$-dimensional classical-statistical theory by virtue of dimensional reduction, and the $3$-dimensional non-Abelian gauge theory is known to exhibit an area law in thermal equilibrium~\cite{montvay-muenster}. Our results may be viewed as nonequilibrium generalization of this.   

\section{Discussion}\label{Discussion}

Real-time lattice QCD is a crucial tool for the quantitative understanding of
the nonperturbative thermalization processes in heavy-ion collisions.
Late-time dynamics requires calculation of nonequilibrium dynamics in the
full quantum theory, which is still a formidable task. Important aspects of
the short-time dynamics in the presence of high gluon occupation numbers,
however, may be calculated using the classical-statistical field theory
limit without further approximations. In particular, for a large class of
initial conditions there is no need to model high momentum degrees of
freedom as particles or to restrict the discussion to the very weak coupling
limit.
In this work we have considered classical-statistical SU(2) gauge theory for
anisotropic initial conditions with high occupation numbers in the
transverse plane on a characteristic scale $\sim \Delta$. We observe Weibel
or primary instabilities with growth rates similar to those obtained from
previous treatments using a variety of setups and
approximations~\cite{Mrowczynski:1988dz, isotropization,
WongYangMills, Romatschke:2006nk}. We observe secondary growth rates for higher-momentum modes
reaching substantially larger values. The secondaries are fluctuation
induced and reflect strongly non-linear dynamics.

Two types of gauge invariant observables have been calculated: Firstly the
time evolution of the spatial part of the energy momentum tensor, which is
related to pressure. Its Fourier transform may be considered as a scale
dependent pressure with a characteristic coarse-graining momentum scale. We
observe that plasma instabilities lead to a rapid "bottom-up" isotropization
up to momenta of about $1$GeV for an energy energy density of $\epsilon = 30
\text{GeV/fm}^3$. The characteristic time scale is given by the inverse
growth rate of about $1 - 2$fm/c. Complete isotropization does not occur on
short time scales. Therefore it is important to note that an effectively
isotropic pressure for a low-momentum range of characteristic momenta is a
crucial ingredient for the use of hydrodynamics, and complete isotropization
may not be necessary to account for its successful applications to describe
RHIC data.

The pressure is a quantity which is not sensitive to long-ranged
fluctuations in contrast to the second type of observables we compute: The
nonequilibrium spatial Wilson loop is shown to become isotropic on similar
characteristic time scales as the coarse grained pressure for low momenta.
At this time all large spatial Wilson loops are found to obey an area law,
which is used to compute a non-vanishing scale with the dimension of a
string tension. The generation of
a non-vanishing scale demonstrates the non-perturbative nature of the
classical-statistical limit of the corresponding quantum theory. We have verified that the pressure and growth rates are insensitive to this scale, which is required for weak-coupling descriptions of these quantities. 

The extension to SU(3) pure gauge theory is straightforward, though computationally more expensive, in order to verify that the changes are indeed minor as expected. Since fermions cannot have large occupation numbers, their contributions to the low-momentum dynamics is expected to be small and the results should give a quantitative account of the short-time physics. \\

We thank Adrian Dumitru, Carsten Greiner, Mike Strickland and Zhe Xu for
very interesting discussions, and Mike for re-calculating some of the
results presented here using his setup. This work is supported in part by
the BMBF grant 06DA267, and by the DFG under contract SFB634.

\appendix
\section{Appendix}\label{Appendix}

In this appendix we demonstrate the insensitivity of our results under variations of volume or spatial lattice distance. All simulations were done with $g=1$ and the initial configurations were generated from the
momentum space probability distribution Eq.~\eqref{InitCond-1} using
the same values for the widths and fixing the normalization such that the energy densities agree. 

\begin{figure}[t!]
\begin{center}
\epsfig{file=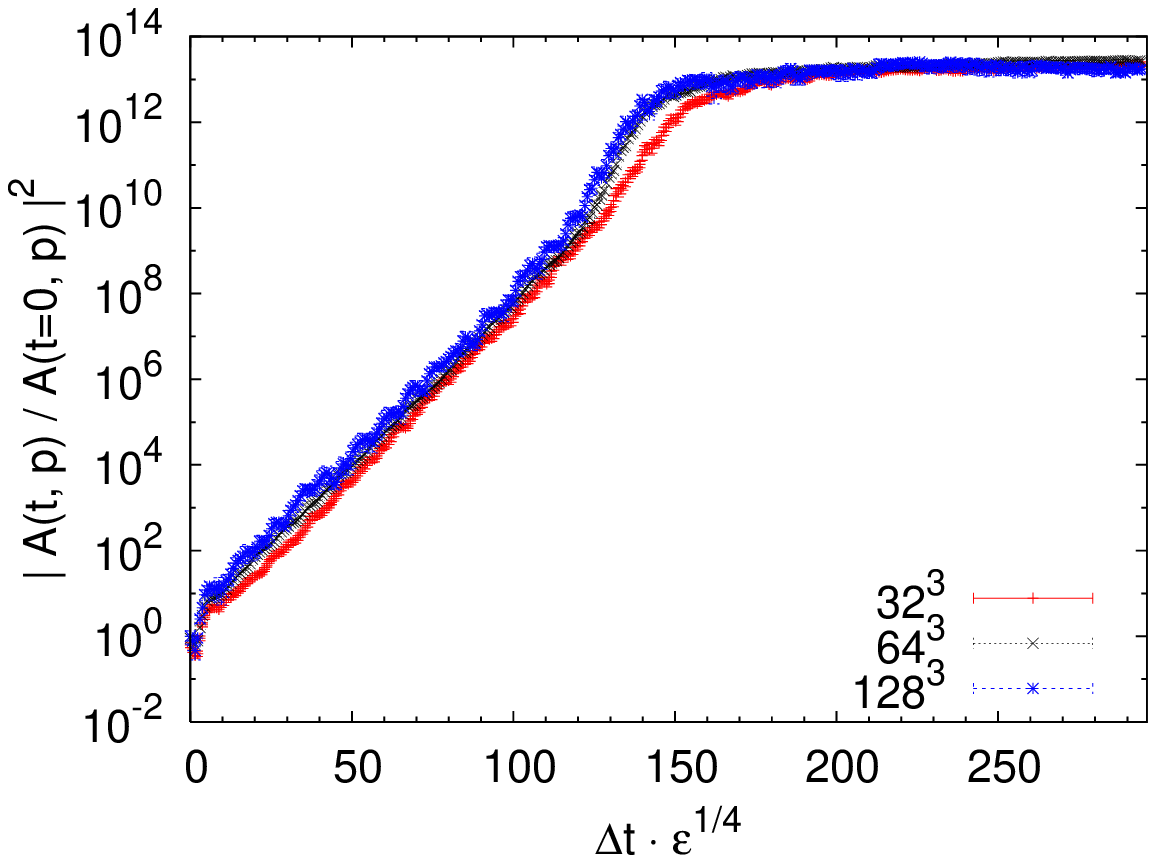, width=6.8cm}
\epsfig{file=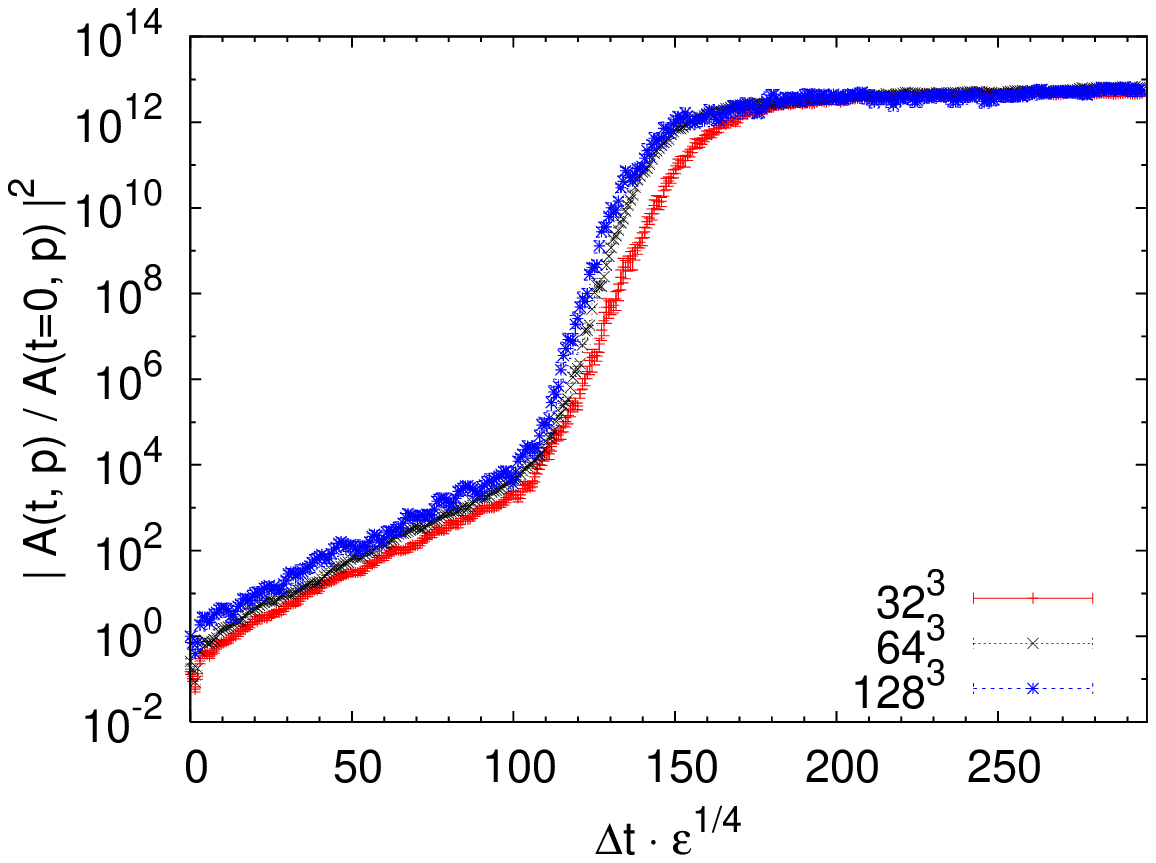, width=6.8cm}
\caption{(Color online) Fourier coefficients of the gauge fields as in Fig.~\ref{A-vs-t} computed for three different lattice volumes. The left panel shows the momentum $p = 0.5 \, \epsf$ while the right panel is for $p = 0.8 \, \epsf$. While the curves computed on 64$^3$- and 128$^3$- lattices coincide well, the one from the 32$^3$- lattice (red) lies a bit lower.}
\label{fig-volume-effects}
\end{center}
\end{figure}

First, we check for possible volume dependences. In Fig.~\ref{fig-volume-effects}
the time evolution of the squared modulus of the gauge field Fourier modes is plotted similarly as in
Fig.~\ref{A-vs-t} as obtained from lattices of
 $32^3 , \, 64^3 \, \text{and} \, 128^3 \, $ sites. On all lattices, we have chosen modes parallel to the z- axis with equal magnitude $p \simeq 0.5 \, \epsf $ and $p \simeq 0.8 \, \epsf $ for the left and the right graph of  Fig.~\ref{fig-volume-effects}, respectively. Volume dependences can hardly be observed for lattices of size 64$^3$ or larger and also on the smallest lattice (32$^3$) only minor deviations occur. 

\begin{figure}[!t]
\begin{center}
\epsfig{file=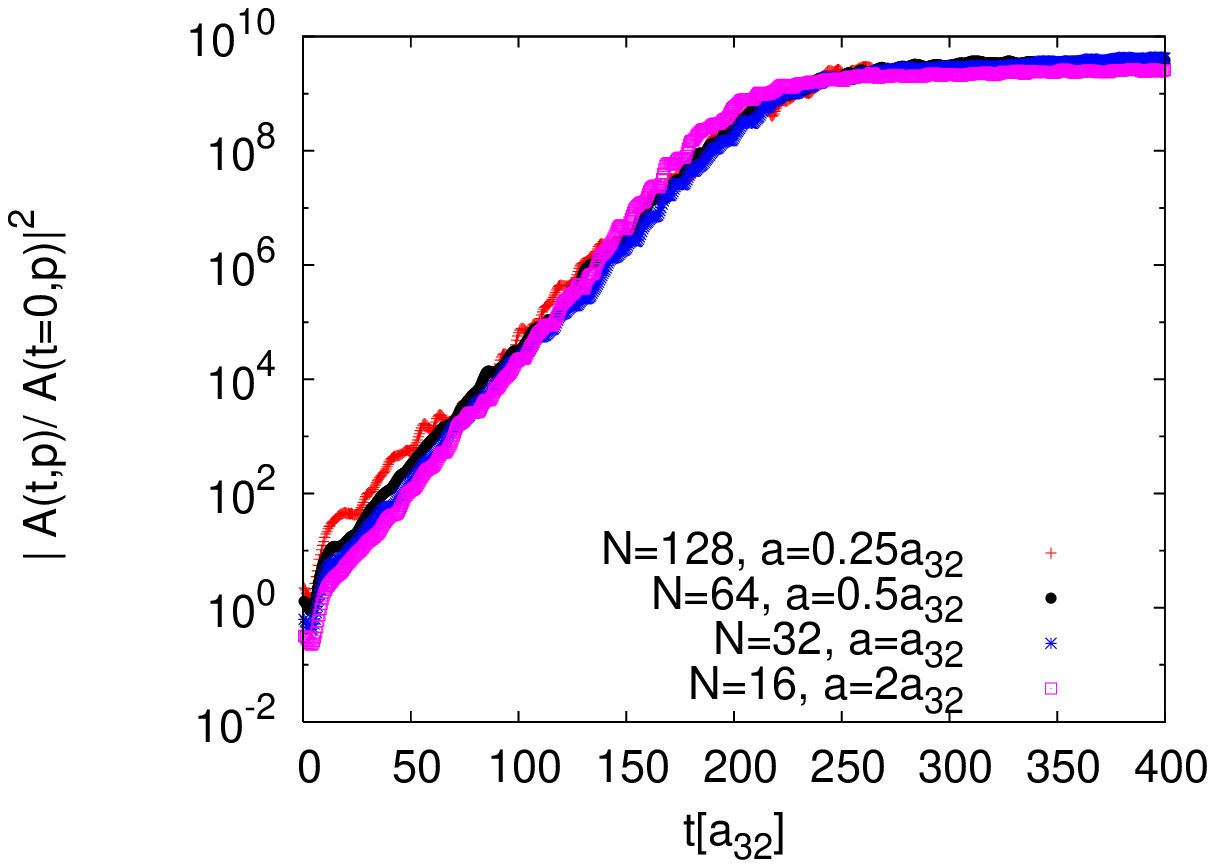, width=6.8cm, angle=0}
\epsfig{file=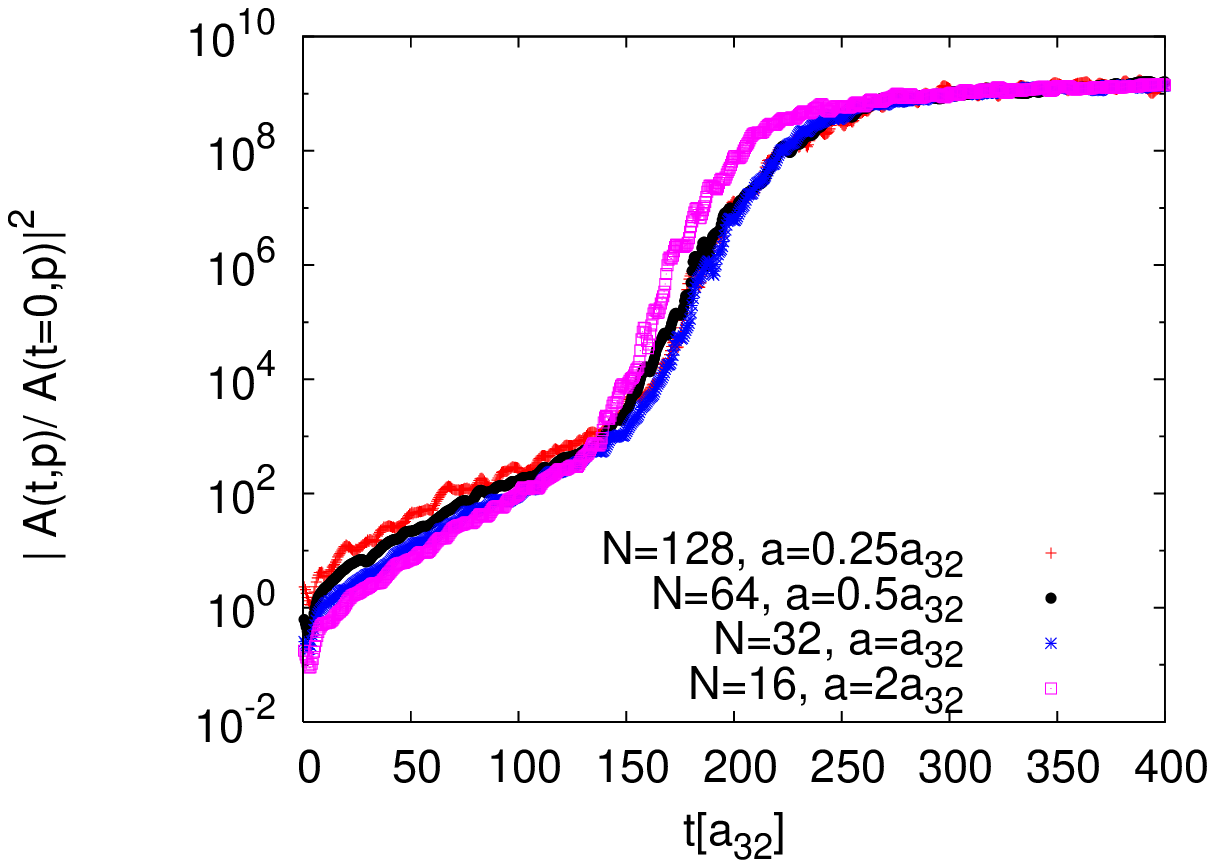, width=6.8cm, angle=0}
\caption{(Color online) The time evolution of the first and second non-zero mode parallel 
to z-axis plotted for systems with different lattice spacings. Time is measured in units of the lattice spacing of 
the $32^3$- lattice. The left graph employs the momentum $p = 0.20 /a_{32}$ while the right graph uses $p =0.39 /a_{32}$. The ordering of the legend corresponds to the ordering of the curves from top to bottom at $t=0$. }
\label{fig-uv-effects}
\end{center}
\end{figure}

Since the UV part of the spectrum is not excited in the beginning, the dependence on the UV cutoff or, equivalently, the lattice spacing should be even milder than the volume dependence. In Fig.~\ref{fig-uv-effects}
the time evolution of Fourier modes is plotted as obtained from simulations on lattices of
different lattice spacings but same volume. We have chosen the first and the second non-zero momentum of the 32$^3$- lattice in the z- direction. The lattice spacings are $ a=0.25, 0.5, 1, 2$ in units of the lattice spacing used for the results presented in the main body of this paper. One observes from Fig.~\ref{fig-uv-effects} that the results are insensitive to UV cutoff changes if the lattice spacing employed in the rest of this paper or smaller is used.

\end{document}